# Classification of FIB/SEM-tomography images for highly porous multiphase materials using random forest classifiers

Markus Osenberg[1, *], André Hilger[1], Matthias Neumann[2], Amalia Wagner[3], Nicole Bohn[3], Joachim R. Binder[3], Volker Schmidt[2], John Banhart[1, 4], Ingo Manke[1]

[1] Institute of Applied Materials, Helmholtz-Zentrum Berlin für Materialien und Energie, Hahn-Meitner-Platz 1, 14109 Berlin, Germany

[2] Institute of Stochastics, Ulm University, Helmholtzstr. 18, 89069 Ulm, Germany

[3] Institute for Applied Materials (IAM-ESS), Karlsruhe Institute of Technology, Hermann-von-Helmholtz-Platz 1, 76344 Eggenstein-Leopoldshafen, Germany

[4] Institute of Materials Science and Technology, Technische Universität Berlin, Hardenbergstr. 36, 10623 Berlin, Germany

* Correspondence: markus.osenberg@helmholtz-berlin.de Tel: +49 30 8062 42079

## Abstract

FIB/SEM tomography represents an indispensable tool for the characterization of three-dimensional nanostructures in battery research and many other fields. However, contrast and 3D classification/reconstruction problems occur in many cases, which strongly limits the applicability of the technique especially on porous materials, like those used for electrode materials in batteries or fuel cells. Distinguishing the different components like active Li storage particles and carbon/binder materials is difficult and often prevents a reliable quantitative analysis of image data, or may even lead to wrong conclusions about structure-property relationships. In this contribution, we present a novel approach for data classification in three-dimensional image data obtained by FIB/SEM tomography and its applications to NMC battery electrode materials. We use two different image signals, namely the signal of the angled SE2 chamber detector and the Inlens detector signal, combine both signals and train a random forest, i.e. a particular machine learning algorithm. We demonstrate that this approach can overcome current limitations of existing techniques suitable for multi-phase measurements and that it allows for quantitative data reconstruction even where current state-of the art techniques fail, or demand for large training sets. This approach may yield as guideline for future research using FIB/SEM tomography.

## Keywords

# 1 Introduction

Unlike naked-eye observations might suggest most natural materials such as rocks, tissue or plants are porous, i.e. have a reduced apparent density and contain voids of various sizes. Pores might allow for mass transport, enlarge the surface area of the material significantly and lead to mechanical properties differing from those of bulk materials. Many functional materials have a complex three-dimensional pore structure that determines their properties and strongly influences their overall functionality. Erosion, degradation, growth or even flexural strength are only a few properties defined by the porosity of materials. This influence of the porous structure on functionality is also found and most often intended in many manufactured materials, for example electrodes, catalyzers, filters or concrete [1-3]. A better understanding of the relationships between the morphology of pore space and the functionality of various materials can help optimizing their functionality. Understanding can be improved by investigating the 3D morphology of the pore space. This is especially important when studying materials in systems for energy storage like battery electrodes where the 3D pore structure is crucial for electrolyte distribution, ion and electron transport and thus the performance of the battery.

A powerful tool for three-dimensional investigations of nano-porous materials is the focused ion beam (FIB) in combination with a high-resolution scanning electron (SEM) or helium-ion microscope. Nearly any material can be investigated, even non-conducting (with helium ions) or fluids (with a cryogenic stage). During FIB/SEM tomography a sample is stepwise sliced by the ion beam while, after every cut, an SEM image is taken of the sliced section. By stacking such images a 3D image of the entire volume – called tomographic reconstruction – is created. However, for porous materials this technique has a major drawback: When taking an SEM image of a sliced sample where pores have been opened, the background of the pore space is illuminated. This background does not belong to the respective slice but to a slice that will appear in later cutting steps. Therefore, this background signal has to be separated from the foreground in the tomographic 3D reconstruction. In order to address this problem, the pore space can be filled with a material like epoxy resin or other [4]. If the sample is very stable (both chemically and mechanically), this technique is well suited and 3D reconstruction of the acquired slices works well. However, the 3D structure of sensitive 3D pore systems may be altered either by the mechanical forces that have to be applied to infiltrate the open volume and the curing process of the infiltrant itself or by chemical reactions with the solvents used.

This may especially be the case for fine agglomerations of light materials like carbon in battery or fuel cell electrodes. This makes filling of such porous materials and the search for suited infiltrants a rather complex field of scientific research. Finally, closed pore systems or pores that are not sufficiently connected to the rest of the pore volume are inaccessible to the infiltrant and cannot be investigated with this method. In such cases, infiltrants have to be avoided. Instead, approaches have to be developed that allow for a separation of foreground and background in FIB/SEM images. This is a quite challenging task and many methods have been proposed recently. The approaches range from pure automatic thresholding (like Otsu [5]), advanced iterative and optic-flow based algorithms that utilize the pore development during milling [6-8], to neural network driven approaches [9-11]. These approaches work well for a specific application only. Most of them either differentiate between two phases (pore and material) [6,10,12], which is suitable for, e.g., concrete, stone or polymer samples, or even for samples with more phases, but with lower accuracy [7], and are thus not suitable for electrode materials comparable to the ones studied in the present paper. Moreover, the workflow required for successful segmentation varies between the different methods described in the literature. In contrast to specialized machine learning approaches such as the mentioned neural networks, the advanced iterative or optic-flow based approaches by Salzer et al. [6] and Moroni and Thiele [7] – from now on called "z-gradient based approaches" – do work without prior training but require an adjustment of the underlying parameters. Furthermore, the ability to deal with additional artefacts caused by the open-pore nature of the sample varies between the different approaches. Neither the gradient-based approaches nor trained machine learning models utilizing simulated data sets can deal with re-deposited material. Depending on the size of the pores within areas of milled material, some of the material removed will redeposit in the pores and result in increasing amounts of additional material over the time. Thus, only training on real data can deal with these measurement-induced morphologic changes. Currently, no simulation-based trainings does account for such artefacts.

The present paper proposes and evaluates a new approach that exploits information gained from multiple signals during ion milling to train a machine learning algorithm, which in turn allows for an improved FIB/SEM-dataset semantic segmentation and then classification of multiple phases – from now on called classification. This implies the advantage of dealing additionally with artefacts like curtaining effects or image gradients simultaneously and even enables multiphase reconstruction without the need of Raw-data preprocessing. Several methods for pore segmentation in images of a lithium ion battery cathode material are compared and an optimized approach is proposed. The investigated cathode consists of highly structured nickel manganese cobalt (NMC) active material with a carbon-based conducting matrix. Counting the voids as a phase implies three different phases.

# 2 Experimental
## 2.1 Materials

The hierarchically structured cathode material investigated in this study was manufactured with enhanced power and energy density in mind [13-17]. The active material was obtained by grinding, spray drying and calcination of pristine $Li(Ni_{1/3}Mn_{1/3}Co_{1/3})O_2$ powder (NM-3100,Toda America) [18]. The resulting NMC particles had a mean particle size of 11.8 µm, a BET surface of 2.52 m$^2$/g and an internal porosity of 40.6%. The carbon-based binder matrix consisted of a polyvinylidene difluoride (PVDF) binder (Solef5130, Solvay Solexis), carbon black (Super C65, Imerys Graphite & Carbon), graphite (KS6L Imerys Graphite & Carbon). These two components were dispersed in N-methyl-2-pyrrolidone (Sigma Aldrich) resulting in a slurry with about 87 wt% NMC, 5 wt% graphite, 4 wt% carbon black and 4 wt% PVDF binder. The slurry was cast on a 20-µm thick aluminum foil and dried. The final active material loading of the electrode was 24 mg/cm².

## 2.2 Methods
### 2.2.1 FIB/SEM tomography

A 3x3 mm² area of the NMC cathode was cut out and fixed on a standard (12.5 mm Ø) aluminum pin stub using conductive silver. After a resting time of 8 h, the sample was transferred into a ZEISS Crossbeam 340. This dual beam machine combines a gallium liquid metal ion source for sample manipulation (in our case milling), a field emission scanning electron microscope (FE-SEM) with GEMINI® optics for high-resolution imaging and a multi-channel gas injection system (GIS) for material deposition.

A 0.5-µm thick protective layer of platinum of 15x15 µm² area was deposited onto the sample in the region of interest using the GIS. The platinum was heated up to 50 °C. In the first 10 minutes, the platinum precursor was reduced to metallic platinum only by the electron beam set to a current of 250 pA at 5 keV. Then, the procedure was continued using a gallium current of 300 pA at 30 keV for 30 minutes. For tomography, the gallium current was set to 700 pA at 30 keV. The plane spacing was 10 nm over an area of 13x13 µm² resulting in 1300 slices and a total dwell time of 3 h.

High-resolution imaging was done using the InLens and the SE2 detector located off to the side in the microscope chamber – from now on called InLens detector and SE2 detector – simultaneously. The electron current was set to 250 pA at 1 keV. Image acquisition took 21 s per frame resulting in a total measure time of 11 h.

Drift correction was applied to the obtained raw data focusing on shine-through artefacts (pore-background). Considering the coordinates X as the width and Y as the height of each SEM image, then Z can be defined as the depth of the measurement. When examining the image stack in the Z-Y direction (the image stack re-slice) that drift correction results in horizontal structures in background areas.

After the drift correction, tomographic image data was de-noised using a non-local means filter with a photometric distance equal to the standard deviation of homogeneous sample areas. The described computations for image processing have been performed using the open source software Fiji [19].

### 2.2.2 Classification testing

Six different classification approaches detecting the three classes: voids (back ground and pores), active material (NMC) and carbon-binder matrix were evaluated. As we want to reduce the amount of additional processing required for the final result we perform no further image processing. Although the selected threshold-based methods in particular would benefit from further image preprocessing, such as advanced background correction, these processing steps were omitted to keep the processing effort uniform.

For the comparison of the different classifications, one slice (#530) of the tomography was completely labeled by hand for each pixel. This manual classification, serving as a ground truth, is used as test data. The machine learning based algorithms tested utilize image information from the vicinity of the voxel to be determined. To avoid an overlap of the test region with the training region, a distance in Z-direction of 300 nm (30 slices) was chosen. This means that no training was performed on slices 501-560. For training, the open source software Ilastik [20] was used (version 1.2.2 in the debug mode). The training data was generated within the Ilastik user interface by manual labelling. However, in contrast to the densely-labeled test slice #530 the training annotation was very sparse.

In order to validate the goodness-of-fit for each classification approach, the intersection over union (IoU) [21] was calculated for each class ($i$), namely for voids, NMC, and the carbon-binder matrix. The IoU was calculated on the full field of view and also just on a small region in the center of the NMC particle in the center of the tomographed volume where no carbon was present. The investigated areas and a cross section of the particle in question are shown in Fig. 1. With that, we wanted to compare the more classical challenge of a binarization between the milled single phase surface (here NMC) and pores, exhibiting low size variations, and material with a multiphase classification, where a large variety of pores and different materials occur. From the $\text{IoU}_i$ for each class we then calculated an average weighted error (awe)

$$\text{awe} = \left(1 - \sum_{i=1}^{n} \text{IoU}_i \cdot p_i \right) \cdot 100\ \% \tag{1}$$

for each of the classification approaches where $p_i$ represents the volume fraction of class $i$. We chose to use the weighted over the non-weighted error to take volume differences per class into account and avoid an overestimation of errors on small classes.

We chose not to use simulations for the evaluation of classification because of the complexity of the multiphase FIB/SEM tomography measurements. We wanted artefacts like curtaining, gradients and re-depositioning to be included in our datasets and there exists no model tool allowing for a simultaneous simulation of all these artefacts.

# 3   Results of image classification

Fig. 1 shows a comparison of the tested classification approaches expressed by the resulting average weighted errors from Eq. 1. The different methods will now be further motivated and explained. For the methods utilizing data from one detector, the SE2 signal was used because this signal yields more relevant data for classification as shown in Fig. 3.

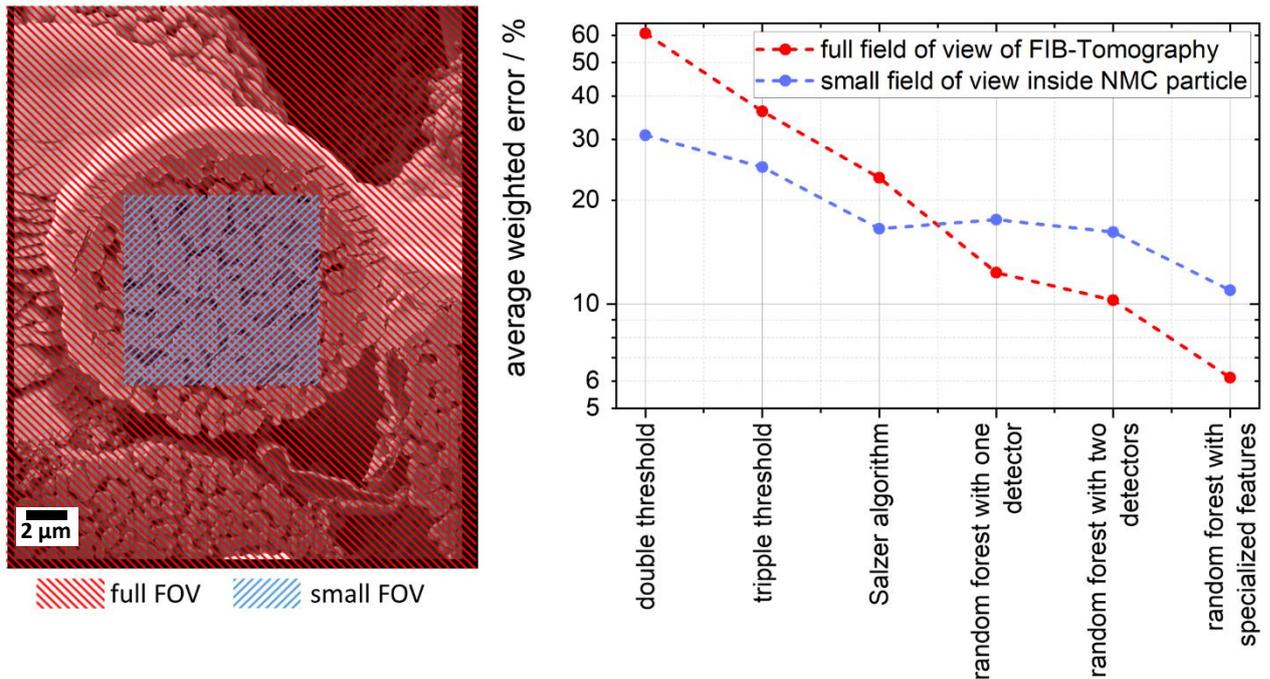

*Fig. 1: Accuracy comparison of the different classification approaches tested, sorted by their full field-of-view averaged weighted error (awe). Slice number 530 is shown on the left. The full and the reduced investigated field of view are marked.*

## 3.1   Double-threshold based

The most straight-forward approach of thresholding a three-phase dataset is a combination of two thresholds. We applied two thresholds based on the carbon peak of the SE2-dataset histogram shown in Fig. 2 (at the gray values 34 and 55, respectively). Voxels with greyscale values below 34 are classified as pores/void, those between 34 and 55 are classified as carbon binder domain, and values above 55 are interpreted as NMC. A weighted error of 60.8 % was accomplished for the whole field of view. The calculated weighted error inside the cropped field of view containing only NMC material and small pores was 30.8 %. Thresholds like the Otsu threshold were ruled out from the beginning as they are designed for binarization and are therefore not suitable for three-phase classification. Local approaches did not work either, as they cannot deal with large variations of pore sizes or large textural variety inside pores.

## 3.2 Triple-threshold based

In Fig. 2, an exemplary histogram of slice number 530 is shown. Based on the manually labeled classes (the ground truth shown in Fig. 4 a), the histogram is separated into the representing gray values per class. The peaks marked 2 and 3 in Fig. 2 show that a third threshold has to be considered to take bright parts of the background into account, especially when not using the InLens detector. The new threshold was set accordingly: 0-27 representing void, 28-48 representing carbon, 49-76 representing NMC and 76-255 representing void again. The thresholds were determined by reducing the information entropy [22] per class to its minimum. However, to find the optimal thresholds for the slice to be tested: slice #530, the entropy was not based on the histogram of the complete 3D image [23,24] but on the labels of the ground truth. The thresholds were calculated by training a decision tree with a depth of two, using the ground truth as a label and only the SE2 signal image as an input. A weighted error of 30.8 % was achieved, significantly improving the results from Section 3.1. With a weighted error of 24.9 % for the internal region inside the NMC particle, the improvement is smaller compared to Section 3.1. This is presumably due to the optimization for the whole field of view and all three classes. Even though (at least on the slice #530) the results improved, having the histogram of Fig. 2 in mind, it becomes clear that a simple classification using global thresholds is not sufficient to separate the classes with high accuracy. As shown in Fig. 4b, this entropy-based optimization also leads to an unrealistic classification at the boundary between classes such as at the boundary between NMC and pores. This observation is in good accordance with the literature [23], where it has been shown that a classification of three-phase materials by global thresholding is error-prone at the boundary between the phases.

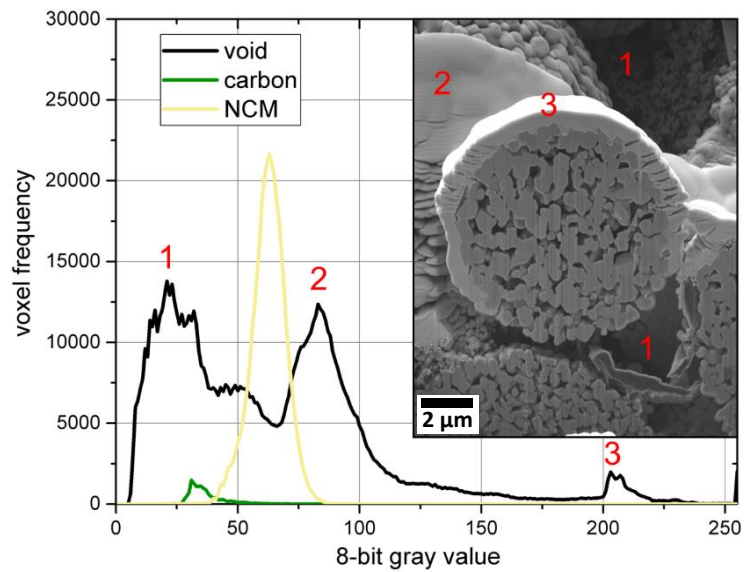

Fig. 2: Histogram of the manually annotated slice (#530) of the FIB/SEM-dataset.

## 3.3 Z-gradient based algorithms

Algorithms such as the one suggested by Salzer et al. [6] do not simply use thresholds to separate different classes. Additionally, they take the pore cutting progress into account. As it compares slices before and after the slice to be classified, it exploits that shine-through artefacts do change gradually during milling until the actual cutting edge is reached. Unfortunately, that approach is designed for binary classification distinguishing between the milling edge and the pore space. It does not work well with re-depositioning, the platinum on top of the sample or multiphase materials in general. This algorithm classifies the void phase in a first step. The remaining area is then separated again using a threshold. Here, the threshold was applied using the InLens-Signal. Each voxel, which has not been classified as void by the z-gradient based algorithm, was either classified as carbon if the InLens-Signal was between 0 and 89 or otherwise, as NMC. With this method, a weighted error of 23.2 % is achieved, again improving overall accuracy. In addition, the error for the reduced field of view is markedly reduced to 16.5 %.

### 3.4 Random forest based classification (one input signal)

The next logical step was to test, whether a machine learning algorithm can outperform any of the previously mentioned approaches. Due to the implementation of 3D filtering for feature generation in Ilastik [20], this method does not only take the pore evolution during milling into account but also deals well with image artefacts in the other dimensions such as image gradients or curtaining. The machine learning algorithms – in our case we choose the random forest classifier – generally work better the more non-correlated input data is used per voxel. Only with the SE2 signal and its 60 standard Ilastik filter varriations called features, the weighted error already dropped to 12.3 %. However, the weighted error for the cropped sub volume inside one NMC particle was 17.6 %, thus larger than the error obtained by the approach based on that of Salzer et al.[6]

### 3.5 Random forest based classification (two input signals)

In a next step, we carried out the same classification including both the SE2 and the InLens signal and the difference of both signals, which leads to a number of 183 features in total. In doing so, the weighted error was reduced to 10.3 %, which is more than two times better than the approach presented in Section 3.3. On the cropped volume, with a weighted error of 16.2 %, the algorithm that was trained for the whole dataset was slightly better than for the approach presented in Section 3.3.

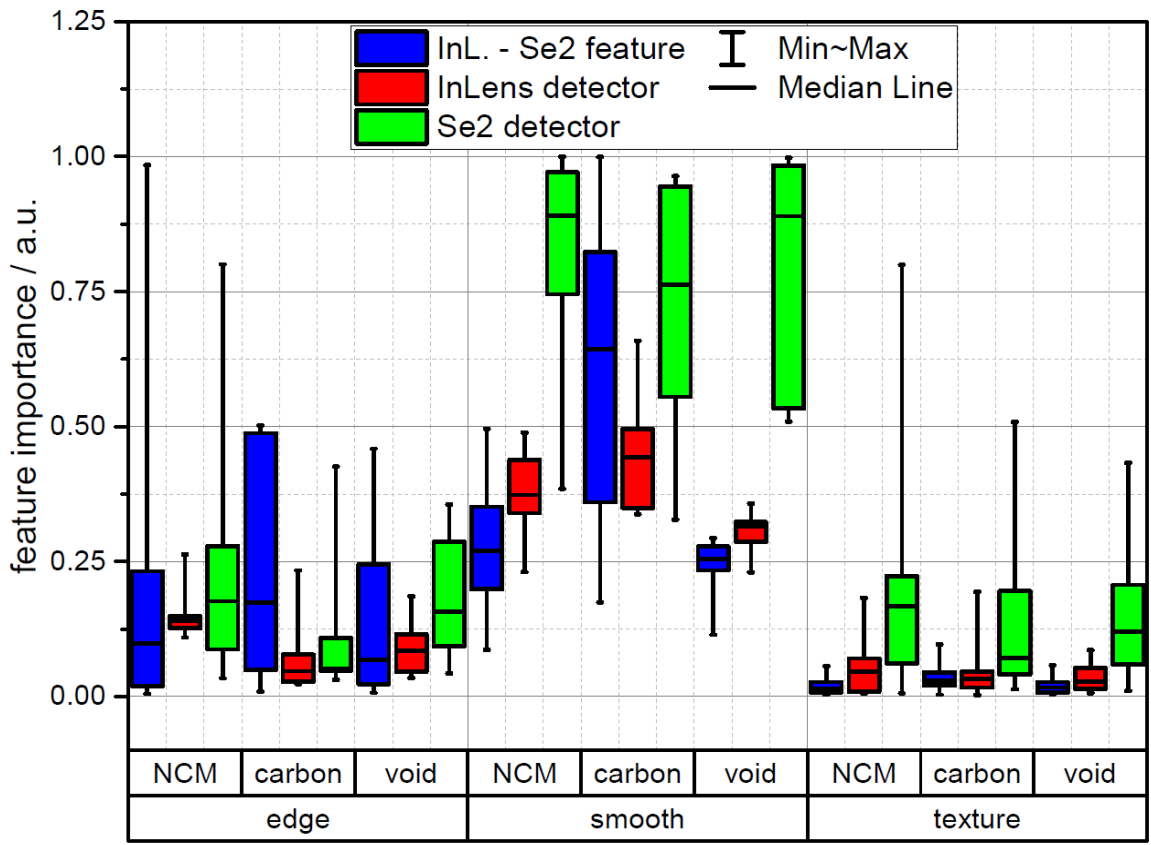

*Fig. 3: Box diagram of the importance of the filter for the random forest classification based on two input signals. The different filters are divided into three classes (edge detecting filters, smoothing filters and texture detecting filters). These filter classes are further separated into their importance concerning the different image classes respectively.*

Fig. 3 shows the resulting importance of different features based on the variable importance [20] in detail. The different filters used are divided into the three categories *edge, smooth,* and *texture*. The group of edge-based filters includes filters that emphasize edges such as derivative-based filters. Smoothing filters are characterized by denoising and blurring of image information. Texture filters such as the Hessian filter help to distinguish between different shapes in the image data. Even though the angled SE2 detector shows a higher relevance for the classification compared to the InLens SE2 detector in the most cases, the subtraction of both is of high importance concerning the classification of carbon. The texture-based filters show an overall low impact on the classification.

## 3.6   Random forest based classification (tailored features)

For the final classification approach, we designed features mainly depending on the evolution of pore space during milling. The idea behind that approach is to make use of the knowledge gained by the method described in Section 3.3 but to let the algorithm learn by itself. Thereby the random forest is trained to interpret the obtained data which turns out to be advantageous compared to the model-based interpretation of the approach in Section 3.3. More precisely, the resulting classification did outperform all the other approaches considered in the present paper by far with a weighted error of 6.1 % on the total data set, which is haft of that of the pure single detector approach (3.4). This result represents another case of a classification algorithm where a hybrid approach combining machine learning with classical image analysis leads to improved results [25].

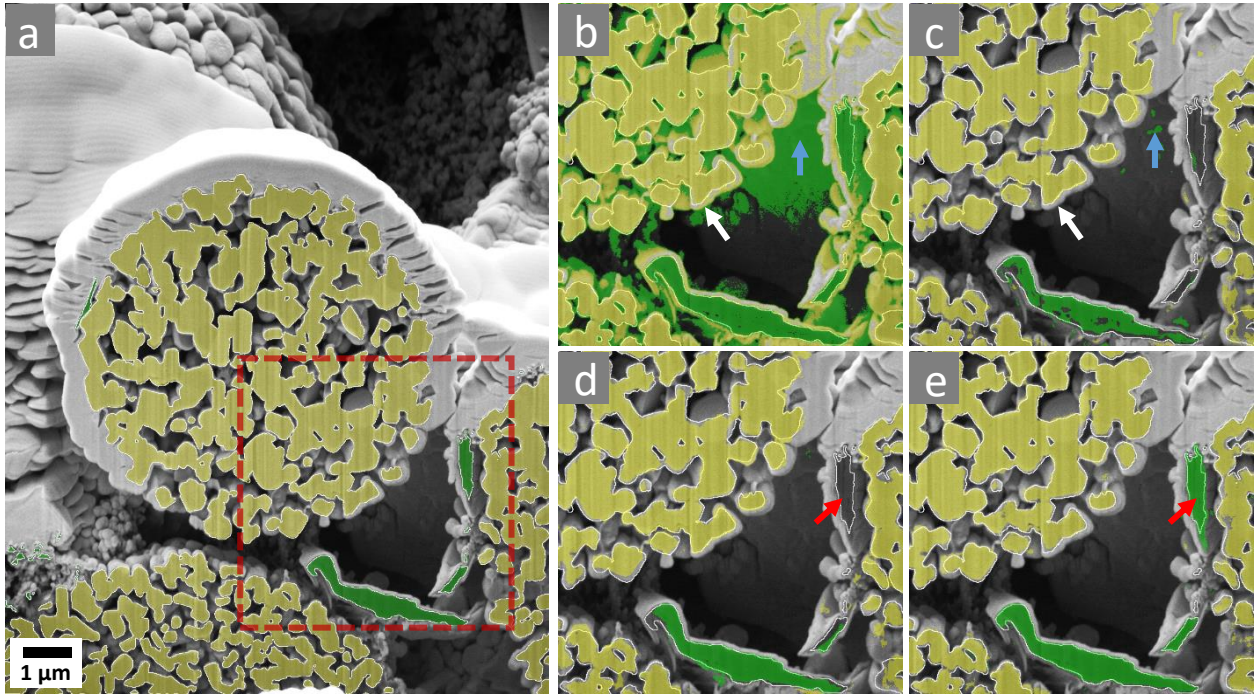

*Fig. 4: Direct comparison of the classification on slice number 530: a) SE2 scan in the middle of the tomography with the ground truth mask (yellow NMC, green carbon matrix), b) Triple-threshold based segmentation (3.2), the white line delimits the border of the ground truth, the white arrow marks wrongly classified redeposited areas, the blue arrow the wrongly classified pore space), c) Random forest segmentation using the angled SE2 signal only (3.4), d) Random forest segmentation using the SE2-signal, the InLens-signal and the difference of both signals (3.5) as an input, the red arrow marks the still wrongly classified carbon phase, e) Random forest segmentation using tailored features taking all detectors and especially the sample development in cutting direction into account (3.6).*

Fig. 4 e) shows detailed results of this classification approach in comparison to the manually segmented test data as well as to other approaches described in Sections 3.2 to 3.6. The region where re-deposition and platinum deposition accrued was well classified. Note that the carbon material can also be classified. No curtaining correction was done but the classifier was also able to compensate for that. Next to the large size variation of the pore areas, the carbon material is also very sparse and varies in size. We found that this was the most challenging aspect of the classification and only the random forest trained on the tailored feature set was able to classify these structures.

With a weighted error of 11.0 %, this approach also outperformed the other approaches in the cropped volume within a single NMC particle. Again, the accuracy is inferior to the overall accuracy because the algorithm was trained and specialized on classifying three phases with a large variety in pore sizes.

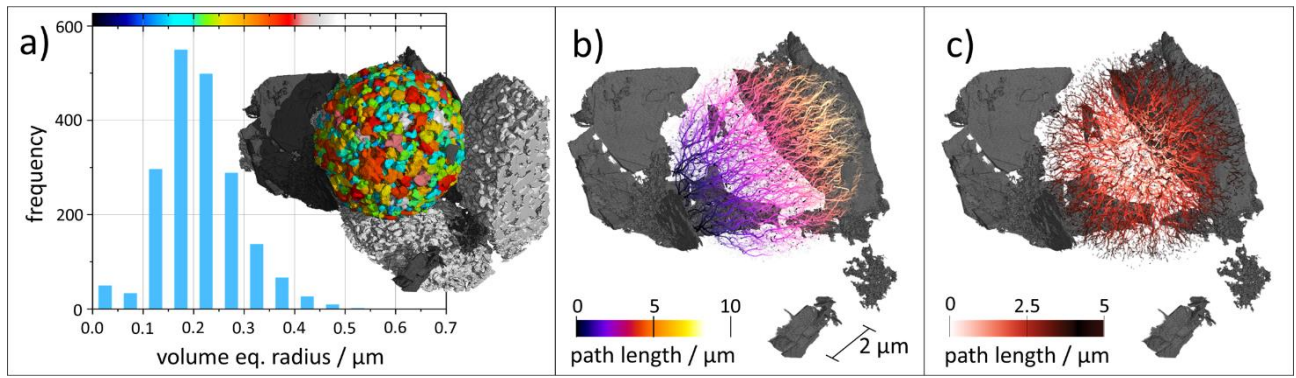

*Figure 5: Renders of the 3D classification: a) shows the particle size distribution, b) shows the electron path through the NMC particle whereas the transparency represents the current density and the color the path length and c) shows the ionic path through the particle pores whereas the transparency represents the current density and the color the path length.*

A precise multiphase classification of porous materials allows for a large variety of analyses. Figure 5 shows an analysis of the size distribution of the primary NMC particles, some paths of the electrons as they are conducted through the NMC material starting from the carbon-connected surfaces and some ion paths through the pores of the material. The trajectories of the electrons represent the shortest path through the NMC phase, starting from uniformly distributed starting points within the NMC to geodesic closest contact point of the NMC and carbon. Since the tomographed particle was on top of the electrode material foil, only one side of the entire particle was connected to the electron conducting carbon matrix. Thus, the lengths of the electron paths are partly larger than the average diameter of the aggregated NMC particle, which is 6.7 µm. On the other side, the path analysis underlines the effectiveness of the particle's porous structure. The inner pores of the particle, meaning the pores enveloped by the particle's convex hull, all can be reached directly without the need of diffusion through the active bulk material as is described in more detail in Section 3.7. The 3D rendering shown in Fig. 5 was done using VG Studio, Paraview [26] and Blender 3D. The analysis of pathways and the segmentation of primary particles was performed using Fiji/morpholibJ [27]. Supplementary video 1 features a full revolution around the volume and shows the fine nature of the carbon matrix. It is obvious that post processing like open-filtering and close-filtering would be rather harmful for the final result.

### 3.7 Further statistical analysis of segmented image data

In this section, we present a further statistical analysis of the image data based on its semantic segmentation by random forests as described in Section 3.6. The focus is on volume fraction of the solid phase as well as on characteristics related to the shortest transportation paths and bottleneck effects in the pore space. The analysis in this section is based on the individual NMC particle, which has been resolved completely by tomographic imaging, *i.e.*, the particle from which the cutout for the small FOV is taken, see Fig. 1.

#### 3.7.1 Volume fraction of the solid phase

To compute the volume fraction of the solid phase of the considered NMC particle – in the following volume fraction – we determine the inner pore voxels by means of the so-called rolling ball algorithm. This algorithm has also been used to determine the inner pores of battery electrodes [28] and paper-based materials [29]. This means that all voxels of the void space are classified as inner pore voxels of the particle, which cannot be reached by a ball intruding from outside with a predefined radius $r$. Here we choose $r$ = 0.15 µm. This gives reasonable results compared to what would visually be determined as inner pores Having determined the inner pore space, we obtain a global NMC volume fraction of the particle of 0.66.

A more detailed analysis reveals a gradient of the local volume fraction. The blue line in Fig. 6 a) shows the local volume fraction as a function of the distance to the particle boundary. More precisely, we compute the local volume fraction for all subvolumes that include voxels having a distance to the boundary between $\delta$ and $\delta$ + 10 nm, for $\delta$ = 0 nm, …, 2840 nm, in steps of 10 nm. The resulting local volume fractions (blue curve) is higher near the boundary of the particle. Note that the uncertainty of the local volume fraction increases with the distance to the boundary due to a decreasing considered volume of the corresponding subvolumes. The red line in Fig. 6 a) shows a cumulative version of the local volume fraction. For a given distance $\delta$ to the boundary, the red line at position $\delta$ represents the volume fraction of all voxels having a distance of less than $\delta$ to the boundary. This red curve indicates that besides the gradient at the boundary of the particle, no further spatial gradients of the volume fraction are present in the considered NCM particle. Moreover, it shows that—as stated above—deviations observed for large distances to the boundary in the blue curve do not affect the global volume fraction of the solid phase.

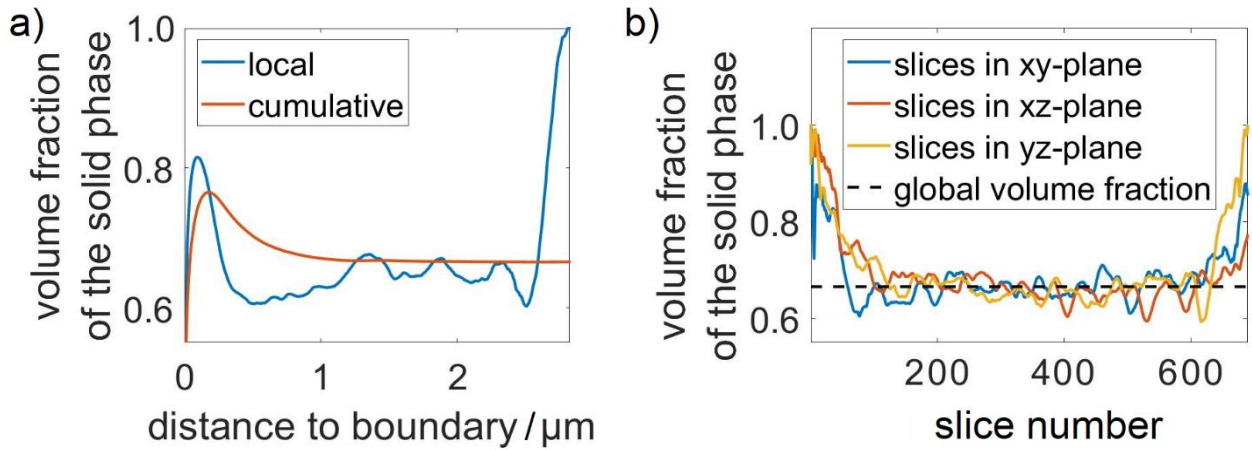

*Fig. 6: (a) Volume fraction depending on the distance to the boundary. At position x, the blue curve represents the volume fraction of all voxels with a distance between $\delta$ and $\delta$+ 10 nm to the boundary and the red curve represents the volume fraction of all voxels with a distance of less than $\delta$ to the boundary. (b) Volume fractions estimated from 2D cross-sections compared to global volume fraction. Cross-sections in xy-, xz-, and yz-plane are considered.*

Under certain assumptions on the homogeneity of the microstructure, it is possible to estimate the volume fraction based on 2D cross-sections, see, e.g., Ch. 10 in Chiu et al. [30]. These assumptions are not fulfilled here due to the gradient in the volume fraction. Even though some generalizations are available in the literature to describe microstructures with structural gradients [31], in our case the following problem occurs: When observing the 2D cross-section of an NMC particle we cannot determine the distance to the boundary in 3D and can thus not deal with this particular gradient. Fig. 6 b) shows the volume fractions estimated from 2D cross-sections, where cross-sections in the xy-, xz-, yz-planes are considered. The volume fraction is overestimated if the cross-section is too far away from the centroid[*] of the particle. Nevertheless, when considering the mean values of the estimated values between slices 150 and 550, we obtain a volume fraction of 0.66 for all cross-sections in xy-, xz-, yz-plane, which coincides with the global volume fraction. The corresponding standard deviation of 0.02 does also not depend on the cross-section. This implies that the volume fraction can be reliably estimated from 2D image data, if the considered slice is sufficiently close to the centroid of the particle. To determine what is sufficient 3D information is necessary.

### 3.7.2 Characteristics related to the length of transportation paths and bottleneck effects

Besides volume fraction, the descriptors "mean geodesic tortuosity" – measuring the length of transportation paths – and "constrictivity" – measuring bottleneck effects – are of crucial importance for the characterization of microstructures, in which transport processes take place [32,33]. The computation of these quantities requires 3D image data, cannot be estimated from 2D cross-sections and is very sensitive to the quality of segmentation. In this section, we quantify the length of transportation paths and bottleneck effects based on the concept of mean geodesic tortuosity and constrictivity, but slightly adapted to the spherical shape of the NMC particles.

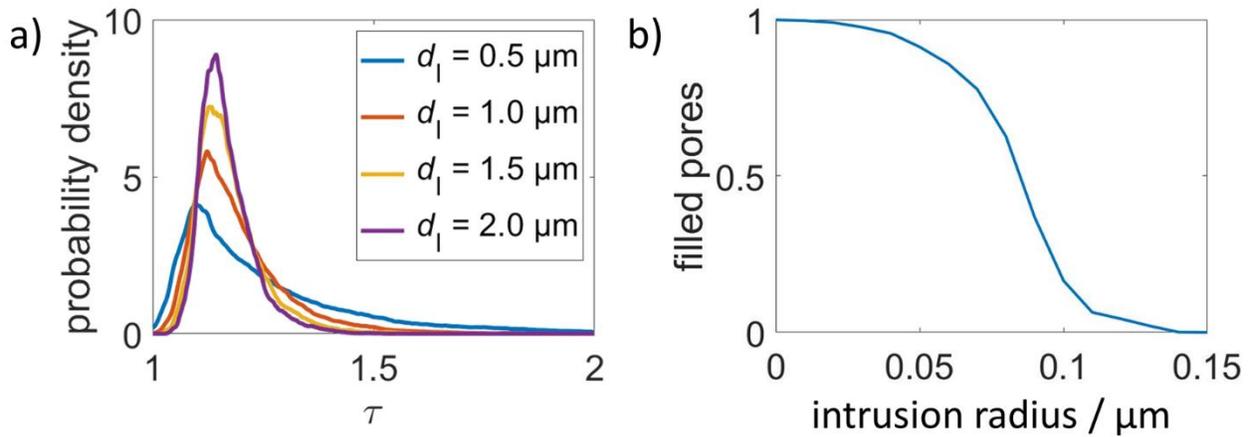

Fig. 7: (a) Distribution of geodesic tortuosity τ for different values of $d_l$. (b)Volume fraction that can be filled by an intrusion of spheres with a predefined intrusion radius r.

For the quantification of the length of transportation paths, we compute a geodesic tortuosity τ for each inner pore voxel located at the boundary of the NMC particle. This means that we compute the length of the shortest path starting at the considered voxel and going through the pore space to the set of those voxels with a distance $d_l$ to the boundary. Then, we define τ as the ratio of the obtained path length and

---

[*] We write centroid here, since the NMC particles are not perfect spheres. In case of a perfect sphere, the centroid would coincide with the midpoint.

$d_l$. For given $d_l$, we obtain a distribution for τ, see

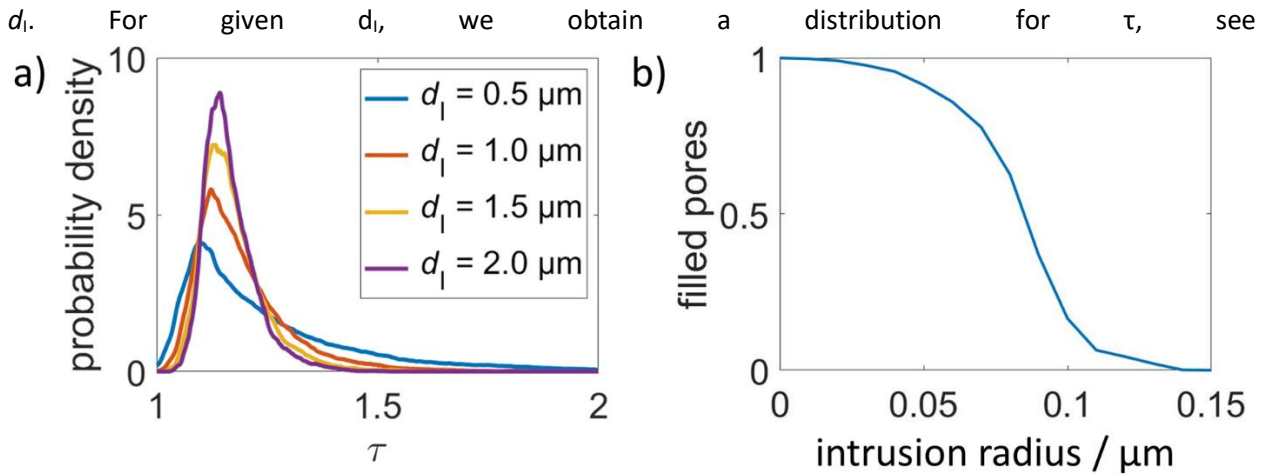

Fig. 7 a).

The mean values (and corresponding standard deviations in brackets) of τ are 1.26 (±0.22), 1.19 (±0.11), 1.17 (±0.07) and 1.17 (±0.06) for $d_l$ = 0.5 µm, 1.0 µm, 1.5 µm and 2.0 µm, respectively. The decrease in the mean values of τ can be explained by the gradient in volume fraction of the solid phase. Close to the boundary there are less pores, the paths can go through. Thus, for smaller $d_l$, the paths become more tortuous, which leads to larger mean values of τ. The decreasing standard deviations with increasing $d_l$ underline the observation in

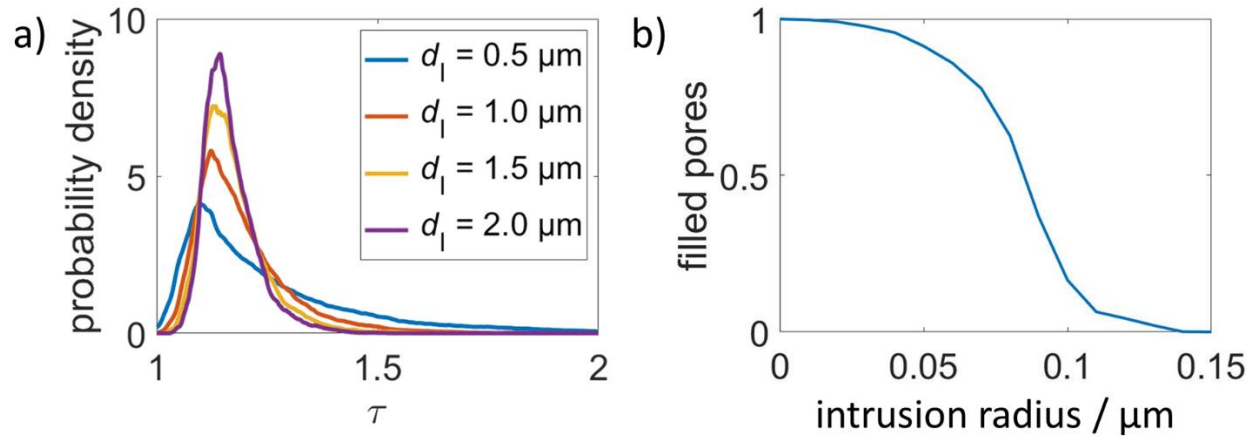

Fig. 7a) that for small $d_l$, more outliers of rather tortuous paths are observed. This effect becomes less pronounced for increasing values of $d_l$.

In order to measure bottleneck effects, we compute the amount of pores that can be filled by an intrusion of spheres with a predefined radius r, where intrusion starts at the boundary of the particle, see

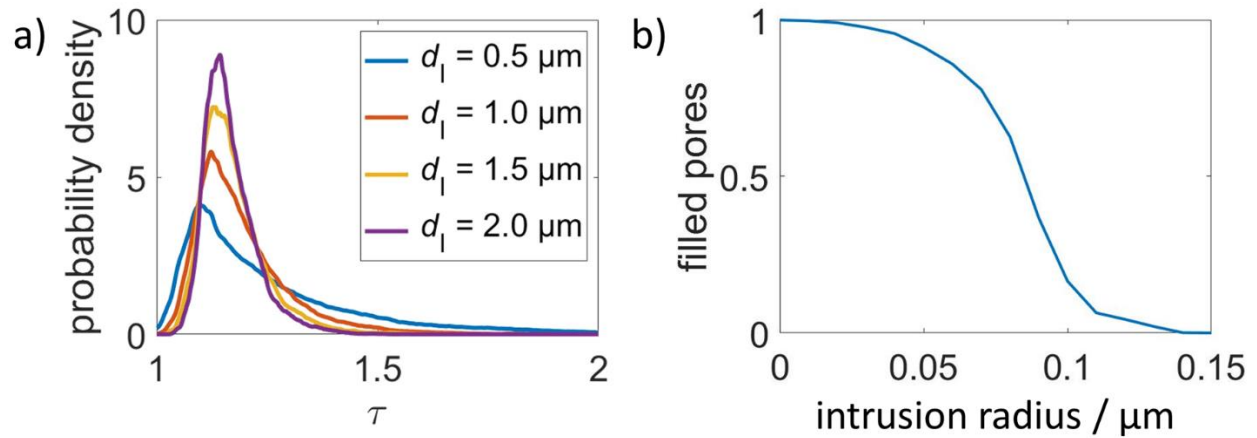

Fig. 7 b). This is the same concept as used for the determination of constrictivity [34]. Doing so, it is possible to see that the microstructure within the NMC particles exhibits a characteristic bottleneck between 0.08 µm and 0.10 µm. While 63 % of the pores can be filled by an intrusion of spheres with radius 0.08 µm, only 17 % can be filled if the radius is increased to 0.10 µm.

## 4 Discussion

We found that the machine learning approach outperforms all the traditional approaches for analyzing complex data. For simple classification tasks as the segmentation of a region inside an aggregated NMC particle, however, machine learning might not be the best choice as classical gradient-based approaches can achieve comparable results without the need for time consuming training. In particular, this holds true, as the results can be improved for those regions by merely focusing on binarization tasks and by applying further pre-processing. For more difficult classification tasks, machine learning approaches are more robust when dealing with artefacts and multiple phases and, in addition, they identify pores with high accuracy. By using machine learning for the classification of 3D FIB/SEM data, the main drawback of non-infiltrated measurements – the lack of finale classification accuracy – becomes manageable. In addition, machine learning approaches are also able to incorporate and learn information that is not directly encoded in the measured gray values but lies in further a priori information such as shapes or arrangements. Without infiltration, much less pre- and post-processing has to be done and the desired region of interest on the sample can be found much easier. As a result, the time per sample at the instrument is reduced, thus allowing for a higher yield of FIB/SEM. In addition, sample alteration e.g. deformations caused by the infiltration and the hardening process can be ruled out and materials that were not accessible due to chemical or contrast reasons are now detectable. The only materials that do not eligible for non-embedded measurements are samples that need the structural support of an embedding material during milling in order to avoid crumbling. For such samples, the approach discussed in the present paper is not applicable.

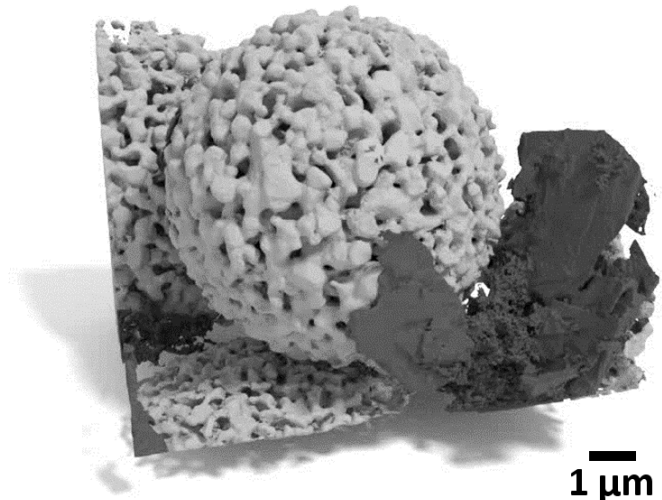

Fig. 5: 3D render of the sample shown in Fig. 4.

Without infiltration, a sample can be inserted directly into the FIB/SEM after fixation. In some cases there is not even the necessity of pre-milling material to reach the desired location. After image acquisition, rather few filtering and image cleaning is required, and the same applies after the classification. In the case of random forest classifiers, training and classifying can be computationally demanding, but for similar datasets, the classifiers can be reused and no further training is required. Except for the data acquisition software, use of software can be limited to open access sources (ImageJ/Fiji [19], Ilastik [20] and Blender [35]/Paraview [26]). These software packages are readily available, well documented and easy to use.

## 5 Conclusions

In the present paper, a random forest classifier was used for the classification of structures in FIB/SEM tomography images of porous multi-phase materials. We show that the technique is capable to take non-infiltrated pores of different sizes into account while dealing with several material phases and identifying re-depositioning artefacts occurring especially in larger pores. We compare the accuracy of different random forest approaches with current state-of-the art techniques. We found that machine-learning based approaches outperform traditional approaches. Furthermore, we show that with specialy filtered input for the random forest and the use of two different detector signals, i.e. the InLens detector and the chamber SE2 detector, the accuracy of the classification can be further improved. Finally, we perform a detailed morphological analysis of the classification involving extracting gradient density, gradient tortuosity, gradient constrictivity and porosity information.

## 6 Acknowledgments

This work has been supported by the German Federal Ministry for Economic Affairs and Energy (BMWi) and granted through Project Management Jülich (03ET6095A, 03ET6095B, and 03ET6095E). The work of MN has been partially funded by the POLiS Cluster of Excellence (EXC 2154/1). This work contributes to the research performed at the Center for Electrochemical Energy Storage (CELEST) Ulm-Karlsruhe.

## 7 References

1  McLaughlin, D., Bierling, M., Moroni, R., Vogl, C., Schmid, G. & Thiele, S. Tomographic reconstruction and analysis of a silver $CO_2$ reduction cathode. *Adv Energy Mater* **10**, 1614-6832, doi:https://doi.org/10.1002/aenm.202000488 (2020).
2  Schulenburg, H., Schwanitz, B., Linse, N., Scherer, G. G., Wokaun, A., Krbanjevic, J., Grothausmann, R. & Manke, I. 3D imaging of catalyst support corrosion in polymer electrolyte fuel cells. *J Phys Chem C* **115**, 14236-14243, doi:https://doi.org/10.1021/jp203016u (2011).
3  Hegge, F., Sharman, J., Moroni, R., Thiele, S., Zengerle, R., Breitwieser, M. & Vierrath, S. Impact of carbon support corrosion on performance losses in polymer electrolyte membrane fuel cells. *J Electrochem Soc* **166**, F956-F962, doi:https://doi.org/10.1149/2.0611913jes (2019).
4  Ender, M., Joos, J., Carraro, T. & Ivers-Tiffee, E. Quantitative characterization of $LiFePO_4$ cathodes reconstructed by FIB/SEM tomography. *J Electrochem Soc* **159**, A972-A980, doi:https://doi.org/10.1149/2.033207jes (2012).
5  Otsu, N. A threshold selection method from gray-level histograms. *IEEE Transactions on Systems, Man, and Cybernetics* **9**, 62-66, doi:https://doi.org/10.1109/TSMC.1979.4310076 (1979).

6   Salzer, M., Spettl, A., Stenzel, O., Smatt, J. H., Linden, M., Manke, I. & Schmidt, V. A two-stage approach to the segmentation of FIB-SEM images of highly porous materials. *Mater Charact* **69**, 115-126, doi:https://doi.org/10.1016/j.matchar.2012.04.003 (2012).

7   Moroni, R. & Thiele, S. FIB/SEM tomography segmentation by optical flow estimation. *Ultramicroscopy* **219**, doi:https://doi.org/10.1016/j.ultramic.2020.113090 (2020).

8   Prill, T., Schladitz, K., Jeulin, D., Faessel, M. & Wieser, C. Morphological segmentation of FIB-SEM data of highly porous media. *J Microsc* **250**, 77-87, doi:https://doi.org/10.1111/jmi.12021 (2013).

9   Kazak, A., Simonov, K. & Kulikov, V. Machine-learning-assisted segmentation of focused ion beam-scanning electron microscopy images with artifacts for improved void-space characterization of tight reservoir rocks. *Spe J* **26**, 1739-1758, doi:https://doi.org/10.2118/205347-PA (2021).

10  Skarberg, F., Fager, C., Mendoza-Lara, F., Josefson, M., Olsson, E., Loren, N. & Roding, M. Convolutional neural networks for segmentation of FIB-SEM nanotomography data from porous polymer films for controlled drug release. *J Microsc* **283**, 51-63, doi:https://doi.org/10.1111/jmi.13007 (2021).

11  Roldan, D., Redenbach, C., Schladitz, K., Klingele, M. & Godehardt, M. Reconstructing porous structures from FIB-SEM image data: Optimizing sampling scheme and image processing. *Ultramicroscopy* **226**, 113291, doi:https://doi.org/10.1016/j.ultramic.2021.113291 (2021).

12  Roeding, M., Fager, C., Olsson, A., Von Corswant, C., Olsson, E. & Loren, N. Three-dimensional reconstruction of porous polymer films from FIB-SEM nanotomography data using random forests. *J Microsc* **281**, 76-86, doi:https://doi.org/10.1111/jmi.12950 (2021).

13  Chen, Z., Wang, J., Chao, D. L., Baikie, T., Bai, L. Y., Chen, S., Zhao, Y. L., Sum, T. C., Lin, J. Y. & Shen, Z. X. Hierarchical porous LiNi$_{1/3}$Co$_{1/3}$Mn$_{1/3}$O$_2$ nano-/micro spherical cathode material: Minimized cation mixing and improved Li$^+$ mobility for enhanced electrochemical performance. *Sci Rep* **6**, doi:https://doi.org/10.1038/srep25771 (2016).

14  Lin, B., Wen, Z., Gu, Z. & Huang, S. Morphology and electrochemical performance of Li[Ni$_{1/3}$Co$_{1/3}$Mn$_{1/3}$]O$_2$ cathode material by a slurry spray drying method. *J Power Sources* **175**, 564-569, doi:https://doi.org/10.1016/j.jpowsour.2007.09.055 (2008).

15  Li, L., Wang, L., Zhang, X., Xie, M., Wu, F. & Chen, R. J. Structural and electrochemical study of hierarchical LiNi$_{1/3}$Co$_{1/3}$Mn$_{1/3}$O$_2$ cathode material for lithium-ion batteries. *ACS Appl Mater Interfaces* **7**, 21939-21947, doi:https://doi.org/10.1021/acsami.5b06584 (2015).

16  Wagner, A. C., Bohn, N., Gesswein, H., Neumann, M., Osenberg, M., Hilger, A., Manke, I., Schmidt, V. & Binder, J. R. Hierarchical structuring of NMC111-cathode materials in lithium-ion batteries: An in-depth study on the influence of orimary and secondary particle dizes on electrochemical performance. *Acs Appl Energ Mater* **3**, 12565-12574, doi:https://doi.org/10.1021/acsaem.0c02494 (2020).

17  Mueller, M., Schneider, L., Bohn, N., Binder, J. R. & Bauer, W. Effect of nanostructured and open-porous particle morphology on electrode processing and electrochemical performance of Li-ion batteries. *Acs Appl Energ Mater* **4**, 1993-2003, doi:https://doi.org/10.1021/acsaem.0c03187 (2021).

18  Dreizler, A. M., Bohn, N., Gesswein, H., Muller, M., Binder, J. R., Wagner, N. & Friedrich, K. A. Investigation of the influence of nanostructured LiNi$_{0.33}$Co$_{0.33}$Mn$_{0.33}$O$_2$ lithium-ion battery electrodes on performance and aging. *J Electrochem Soc* **165**, A273-A282, doi:https://doi.org/10.1149/2.1061802jes (2018).

19  Schindelin, J., Arganda-Carreras, I., Frise, E., Kaynig, V., Longair, M., Pietzsch, T., Preibisch, S., Rueden, C., Saalfeld, S., Schmid, B., Tinevez, J. Y., White, D. J., Hartenstein, V., Eliceiri, K., Tomancak, P. & Cardona, A. Fiji: an open-source platform for biological-image analysis. *Nat Methods* **9**, 676-682, doi:https://doi.org/10.1038/nmeth.2019 (2012).

20  Berg, S., Kutra, D., Kroeger, T., Straehle, C. N., Kausler, B. X., Haubold, C., Schiegg, M., Ales, J., Beier, T., Rudy, M., Eren, K., Cervantes, J. I., Xu, B. T., Beuttenmueller, F., Wolny, A., Zhang, C., Koethe,

U., Hamprecht, F. A. & Kreshuk, A. Ilastik: interactive machine learning for (bio) image analysis. *Nat Methods* **16**, 1226-1232, doi:https://doi.org/10.1038/s41592-019-0582-9 (2019).
21  Jaccard, P. The distribution of the flora in the alpine zone.1. *New Phytologist* **11**, 37-50, doi: https://doi.org/10.1111/j.1469-8137.1912.tb05611.x (1912).
22  Shannon, C. E. A mathematical theory of communication. *The Bell System Technical Journal* **27**, 379-423, doi: https://doi.org/10.1002/j.1538-7305.1948.tb01338.x (1948).
23  Schlueter, S., Sheppard, A., Brown, K. & Wildenschild, D. Image processing of multiphase images obtained via X-ray microtomography: A review. *Water Resour Res* **50**, 3615-3639, doi:https://doi.org/10.1002/2014wr015256 (2014).
24  Kapur, J. N., Sahoo, P. K. & Wong, A. K. C. A new method for gray-level picture thresholding using the entropy of the histogram. *Comput Vision Graph* **29**, 273-285, doi:https://doi.org/10.1016/0734-189x(85)90125-2 (1985).
25  Furat, O., Wang, M. Y., Neumann, M., Petrich, L., Weber, M., Krill, C. E. & Schmidt, V. Machine learning techniques for the segmentation of tomographic image data of functional materials. *Front Mater* **6**, doi:https://doi.org/10.3389/fmats.2019.00145 (2019).
26  Ahrens, J., Geveci, B. & Law, C. ParaView: An end-user tool for large data visualization. *Visualization Handbook* (2005).
27  Legland, D., Arganda-Carreras, I. & Andrey, P. MorphoLibJ: integrated library and plugins for mathematical morphology with ImageJ. *Bioinformatics* **32**, 3532-3534, doi:https://doi.org/10.1093/bioinformatics/btw413 (2016).
28  Kuchler, K., Prifling, B., Schmidt, D., Markoetter, H., Manke, I., Bernthaler, T., Knoblauch, V. & Schmidt, V. Analysis of the 3D microstructure of experimental cathode films for lithium-ion batteries under increasing compaction. *J Microsc* **272**, 96-110, doi:https://doi.org/10.1111/jmi.12749 (2018).
29  Machado Charry, E., Neumann, M., Lahti, J., Schennach, R., Schmidt, V. & Zojer, K. Pore space extraction and characterization of sack paper using µ-CT. *J Microsc* **272**, 35-46, doi:https://doi.org/10.1111/jmi.12730 (2018).
30  Chiu, S. N., Stoyan, D., Kendall, W. S. & Mecke, J. Stereology. *Stochastic Geometry and its Applications*, 411-452, doi:https://doi.org/10.1002/9781118658222.ch10 (2013).
31  Hahn, U., Micheletti, A., Pohlink, R., Stoyan, D. & Wendrock, H. Stereological analysis and modelling of gradient structures. *J Microsc* **195**, 113-124, doi:https://doi.org/10.1046/j.1365-2818.1999.00487.x (1999).
32  Stenzel, O., Pecho, O., Holzer, L., Neumann, M. & Schmidt, V. Big data for microstructure-property relationships: A case study of predicting effective conductivities. *AIChE* **63**, 4224-4232, doi:https://doi.org/10.1002/aic.15757 (2017).
33  Neumann, M., Stenzel, O., Willot, F., Holzer, L. & Schmidt, V. Quantifying the influence of microstructure on effective conductivity and permeability: Virtual materials testing. *Int. J. Solids Struct.* **184**, 211-220, doi:https://doi.org/10.1016/j.ijsolstr.2019.03.028 (2020).
34  Holzer, L., Wiedenmann, D., Münch, B., Keller, L., Prestat, M., Gasser, P., Robertson, I. & Grobéty, B. The influence of constrictivity on the effective transport properties of porous layers in electrolysis and fuel cells. *J Mater Sci* **48**, 2934-2952, doi:https://doi.org/10.1007/s10853-012-6968-z (2013).
35  Community, B. O. Blender - a 3D modelling and rendering package. *Stichting Blender Foundation, Amsterdam* (2018).